\documentclass[11pt]{JHEP3}

\usepackage{amssymb,amsfonts,amsmath}
\usepackage{graphicx}

\newcommand{\be}{\begin{equation}}
\newcommand{\ee}{\end{equation}}

\newcommand{\cl}{\mathcal{L}}

\newcommand{\bi}{\begin{itemize}}
\newcommand{\ei}{\end{itemize}}

\newcommand{\bnum}{\begin{enumerate}}
\newcommand{\enum}{\end{enumerate}}

\newcommand{\bvb}{\begin{verbatim}}

\newcommand{\ba}{\begin{eqnarray}}
\newcommand{\ea}{\end{eqnarray}}

\newcommand{\eq}[1]{(\ref{#1})}

\title{ Notes on holographic Schwinger effect}

\author{
Xing  Wu$^{a,b,c}$
\\

$^a$
{\it Department of Physics, North University of China, \\ 3 Xueyuan Road, Taiyuan, Shanxi 030051, China}\\
$^b$
{\it Key Laboratory of Frontiers in Theoretical Physics, Institute of Theoretical Physics, Chinese Academy of Sciences, \\ P.O. Box 2735, Beijing 100190, China}\\
$^c$
{\it Kavli Institute for Theoretical Physics China, Chinese Academy of Sciences,\\ Beijing 100190, China}\\
E-mail:   \email{xwu@itp.ac.cn}
}

\abstract{
 We use the method of evaluating the decay rate in terms of the imaginary part of a probe brane action to study the holographic Schwinger effect. In the confining D3-branes case, we find that the Schwinger effect occurs at energy scales higher than the Kaluza-Klein mass, indicating the absence of such effect when the dual gauge field theory can be regarded as an 2+1 dimensional theory. This property is independent of the configuration of the probe brane. In the case of D3-branes with a $B$ field dual to a noncommutative super Yang-Mills theory, we study how the decay rate is affected by the noncommutative effect.
}

\begin{document}

\section{Introduction}
The Schwinger effect refers to the process of creation of charged particles out of vacuum in the presence of an external electric field, which is an interesting phenomena in quantum field theory. The particle creation rate was first studied in the context of electron-positron pair production in quantum electrodynamics by Schwinger in \cite{Schwinger:1951nm}, where the production rate\footnote{
Note we will not distinguish the terms ``vacuum decay rate" and ``pair production rate", although it is argued \cite{Cohen:2008wz} that  the two are different in a strict sense.
}
of particles with mass $m$ and charge $e$ in an electric field $E$
is given as
\be\label{Schwinger rate}
w=\frac{(eE)^2}{4\pi^3}\sum_{n=1}^{\infty}\frac{1}{n^2} e^{-n\frac{\pi m^2}{eE}}
\ee
where the coupling is weak and the electric field is not strong either. The finite coupling effect is considered in \cite{Affleck:1981bma}.

Despite abundant theoretical investigations in the literature, there is no direct experimental test of this effect so far. This is because the pair production rate is very small, and to have a sizable effect, the applied electric field is typically required to be of order $10^{18} V/m$, which renders it difficult to realize in usual laboratories. There are also interesting ideas towards possible experimental study of this effect, e.g. using graphene \cite{Allor:2007ei}, or using the cold atom systems \cite{Kasper:2014uaa}.

The Schwinger effect is not confined to QED, and the form of the above formula applies to a more general context. Indeed, it is a generic feature of vacuum instability under external fields, and the created particles can be scalar particles or quark-antiquark pairs, in addition to electron-positron pairs.

The holographic study of the Schwinger effect within the context of the AdS/CFT correspondence\cite{Maldacena:1997re,Witten:1998qj,Gubser:1998bc} was first proposed in \cite{Semenoff:2011ng}, where the authors study the pair creation of the W-bosons in the Coulomb branch of $\mathcal{N}=4$ super-Yang Mills (SYM) theory, which is dual to $N$ coincident D3-branes with a probe D3-brane placed at a finite separation from the source. In this case, they calculated the instanton action as the dominant part of the exponent by evaluating the action of a string ending on the instanton circle on the probe D3-brane. They also found a critical electric field where the radius of the instanton circle shrinks to zero. This value coincides with the critical value of the Dirac-Born-Infeld (DBI) action of the probe D3-brane with an electric field turned on. This proposal is generalized in other cases \cite{Bolognesi:2012gr,Sato:2013pxa,Kawai:2013xya,Fischler:2014ama}.  The key technique of the proposal is similar to that of the holographic calculation of the VEV of a {\it circular} Wilson loop \cite{Maldacena:1998im,Brandhuber:1998bs,Rey:1998bq}. Later on, an potential analysis \cite{Sato:2013iua} was proposed to study the effective potential of the particle pair by evaluating the {\it rectangular} Wislon loop VEV. This method also reveals that the critical value obtained from the DBI action determines the key feature of the effective potential. Further study in this direction can be found in \cite{Sato:2013dwa,Sato:2013hyw,Kawai:2015mha,Fischler:2014ama,Fadafan:2015iwa,Ghodrati:2015rta}.

The aforementioned studies are basically within the Coulomb branch, i.e. a D3-brane probing $N$ D3-branes. Moreover, what has been calculated in \cite{Semenoff:2011ng} and related works is  just the leading exponent corresponding to the on-shell action of the instanton, not the full decay rate. In particular, the part analogous to the factor in front of the summation in \eq{Schwinger rate} is still to be determined. In contrast, the authors of \cite{Hashimoto:2013mua} studied the full decay rate in the $\mathcal{N}=2$ supersymmetric QCD dual to a D7-brane probing $N$ D3-branes, where the D7-brane adds a U(1) flavor to the SYM. In this context, as one turns on a constant electric field on the probe brane, the original vacuum develops an instability, indicated by the DBI action becoming imaginary. Of course the system will reach a final stable state with a steady current of charged particles.  It is proposed that the {\it full} vacuum decay rate is exactly given by the imaginary part of the DBI action. In particular, a key point is that, if one is only concerned with the onset of this instability, not the final state, then one uses the probe brane configuration  as a solution to the embedding equation derived from the DBI action {\it in the absence of the U(1) field}. This corresponds to evaluating the VEV of the  Euler-Heisenberg effective Lagrangian in the initial vacuum without turning on the external field. Studies in this direction can be found in \cite{Hashimoto:2014dza,Hashimoto:2014yya,Ghodrati:2015rta}.

In this paper, we further explore this method of evaluating the imaginary part of the DBI action to obtain the vacuum decay rate. We study two Dp/Dq brane configurations: (1) a probe D7 brane in $N$ D3-branes wrapped around a circle, dual to a confined SYM in 2+1 dimensions; (2) a probe D7 brane in $N$ D3-branes with a nonvanishing Neveu-Schwarz (NS) $B$ field, dual to noncommutative super Yang-Mills (NCSYM).

\section{Probe D7 brane and confining D3-branes}
The key idea of evaluating the vacuum decay rate via the DBI action is the following. The vacuum is given by a Dp/Dq brane configuration which is a solution to the embedding equation arising from the probe brane action {\it without} any world volume gauge field. Then possible instability of the vacuum is caused by turning on the world volume gauge field. Therefore, the original probe brane configuration is in general {\it not a solution} to the embedding equation in the presence of the gauge field. Indeed, we don't need to follow the full evolution of the system since we are only concerned with the onset of the instability induced by the gauge field. Then, as proposed in \cite{Hashimoto:2013mua}, the decay rate is given by the imaginary part of the DBI action of the original probe brane configuration with a U(1) field turned on.

\subsection{Confining D3-branes}

Consider a $\mathcal{N}=4$ SYM with t'Hooft coupling $\lambda=g_{YM}^2N$ in 3+1 dimensions compactified on a circle of finite radius $R_w$ with antiperiodic boundary conditions imposed on the fermions, which make the fermions massive and therefore completely break supersymmetry \cite{Witten:1998qj}. At long distances, or low energies, the theory becomes effectively a 2+1 dimensional confining theory (pure SYM without supersymmetry) with 't Hooft coupling $\lambda_3=\lambda/R_w$. At large $N$ and strong t'Hooft coupling, this can be holographically described by the near-horizon limit of $N$ coincident D3-branes wrapped around a circle of radius $R_w$, given by \cite{Nishioka:2006gr,Klebanov:2007ws}
\be\label{D3 soliton}
ds^2=\frac{r^2}{R^2}(-dt^2+dx_1^2+d x_2^2+fdx_3^2)+\frac{R^2dr^2}{r^2f}+R^2 d\Omega_5^2,
\ee
where $f=1-r_t^4/r^4$, and the compact direction is given by $x_3\equiv w\sim w+\beta_w$, $\beta_w=2\pi R_w$.
To avoid conical singularity at $r=r_t$, the period must satisfy
\be
\beta_w=\frac{R^2 \pi}{r_t}.
\ee
The Kaluza-Klein (KK) mass scale is given by
\be
M_{KK}=\frac{2\pi}{\beta_w}=\frac{2r_t}{R^2}.
\ee
According to the spirit of the KK reduction, at energy scales less than $M_{KK}$, one can neglect the massive KK modes of the compact dimension and obtain a reduced theory with lower dimensions.

To introduce flavors, one can add $N_f$ probe branes \cite{Karch:2002sh} whose backreaction to the D3-brane geometry is ignored, corresponding to taking the quenched limit in lattice gauge theory. Here the probe is one D7-brane ($N_f=1$), whose action is given by the sum of the DBI part and the Chern-Simons part
\be
S=-T_7\int d^{7+1}\sigma \sqrt{-\det(G_{ab}+2\pi\alpha' F_{ab}})+\frac{(2\pi\alpha')^2}{2}T_7\int C^{(4)}\wedge F\wedge F,
\ee
where $G_{ab}$ is the pullback of the metric on the probe brane, while $F$ is the U(1) gauge field on the brane world volume. In the study of the holographic Schwinger effect, the decay rate is determined from the imaginary part of the action, therefore only the DBI part is relevant.

\subsection{Configuration D7$\|$D3}
Consider the configuration
\be
\begin{array}{ccccccccccc}
   & 0 & 1 & 2 & 3 & 4& 5 & 6 & 7 & 8 & 9\\
D3 & \checkmark & \checkmark & \checkmark & \checkmark&  & &  &  & & \\
D7 &\checkmark  & \checkmark & \checkmark& \checkmark  & \checkmark   &\checkmark & \checkmark& \checkmark &   &  \\
\end{array}
\ee
where the probe D7-brane wraps the $S^1$ of $x_3\equiv w$. In addition, the $S^5$ metric can be written as
\be
d\Omega_5^2=d\theta^2+\sin^2\theta d\phi^2+\cos^2\theta d\Omega_3^2,
\ee
where, by symmetry, one can choose $\phi=const$ and $\theta=\theta(r)$. Moreover, we will only consider the massless case where the separation of the probe and the D3-branes is zero, correspondingly $\theta=0$ such that the $S^3$ is the maximal one. Of course, one needs to show that this is indeed a solution to the embedding equation in the absence of the gauge field. Since the embedding equation can be equivalently derived in both Euclidean and Lorentzian action in the static case, and the Euclidean action of the solution \eq{D3 soliton} is the same as that of a black hole, one can use the result of \cite{Mateos:2007vn} directly and see that this is indeed a solution.

 In the static gauge, the induced metric on the D7 brane is
\be
ds_{D7}^2=\frac{r^2}{R^2}(-dt^2+dx_1^2+d x_2^2+fdw^2)+\frac{R^2dr^2}{r^2f}+R^2 d\Omega_3^2.
\ee
Now the DBI action yields
\be
\cl=-T_7 2\pi^2\int_{r_t}^\infty dr r^3\sqrt{1-\frac{r_*^4}{r^4}},
\ee
where $r_*^4\equiv (2\pi\alpha')^2 E^2R^4$ and $\cl\equiv S_{DBI}/V_3 2\pi R_w$, with $V_3$ being the volume of  $t,\vec x_2^2$, and $2\pi^2$ is the volume of the unit $S^3$. The imaginary part is
\be
{\rm Im}\cl=T_7 2\pi^2\int_{r_t}^{r_*}dr r^3\sqrt{\frac{r_*^4}{r^4}-1},
\ee
which can be evaluated as
\be
{\rm Im}\cl=\frac{N}{16\pi^2}E^2\left(\frac{\pi}{2}-\frac{\sqrt{\gamma-1}}
{\gamma}-\arctan\frac{1}{\sqrt{\gamma-1}}\right),
\ee
where $\gamma\equiv r_*^4/r_t^4>1$.  It is easy to see that the expression within the parentheses is always positive for $\gamma>1$. Moreover, as $\gamma\rightarrow\infty$, i.e. $r_t\rightarrow 0$,  corresponding to the non-confining case, the result becomes  ${\rm Im}\cl=\frac{N}{32\pi} E^2$, which coincides exactly with the result obtained in \cite{Hashimoto:2013mua} for the usual D3/D7 configuration.

As $\gamma\rightarrow1$, the critical radius $r_*$ approaches the IR cutoff $r_t$. This implies a lower bound on the electric field,
\be\label{critical E}
E>E_c\equiv \frac{r_t^2}{2\pi\alpha' R^2},
\ee
below which ${\rm Im}\cl=0$ and there is no Schwinger effect. Indeed, in the confining case, the electric field should be large enough to separate the confined quark and antiquark. This lower bound is also consistent with the result from the potential analysis in \cite{Sato:2013dwa, Sato:2013hyw}. Note that there the configuration is a probe D3 brane in the background of $N$ D3-branes, corresponding to the Coulomb branch, where the created particles are the W-bosons of $SU(N+1)\rightarrow SU(N)\times U(1)$. In the case studied here, the configuration is different. Yet the physics of liberating quarks from confinement is the same.

The electric field $E$ has the dimension of mass (energy) squared. The energy scale at which the Schwinger effect occurs is characterized by
\be
\sqrt{E_c}\sim\frac{r_t}{R\sqrt{\alpha'}}=\frac{r_t}{Rl_s},
\ee
compared with the KK mass scale $M_{KK}$,
\be\label{comparison Mkk}
\frac{\sqrt{E_c}}{M_{KK}}\sim \frac{R}{l_s}\sim \lambda^{1/4}\gg1.
\ee
As a consequence, at large $N$ and large $\lambda$, the electric field with $E>E_c$ probes the massive KK modes, which essentially contribute to the Schwinger effect.
In other words, at energy scales below $M_{KK}$ where the strongly coupled large $N$ theory is effectively reduced from 3+1 to 2+1 dimensions, there is no Schwinger effect. This also explains that the form of the dependence of the decay rate on the electric field. In a general $d$ dimensional spacetime, the decay rate is proportional to $E^{d/2}$ \cite{1996PhRvD..53.7162G}. Here we have ${\rm Im}\cl\sim E^2$, instead of $E^{3/2}$, indicating that the Schwinger effect occurs effectively in 3+1 dimensions.

\subsection{Configuration D7-$\overline{ \rm\bf D7}$$\bot$D3}

One may also consider another configuration
\be
\begin{array}{ccccccccccc}
   & 0 & 1 & 2 & 3 & 4& 5 & 6 & 7 & 8 & 9\\
D3 & \checkmark & \checkmark & \checkmark & \checkmark&  & &  &  & & \\
D7 &\checkmark  & \checkmark & \checkmark&  & \checkmark   &\checkmark & \checkmark& \checkmark &  \checkmark  &  \\
\end{array}
\ee
where the probe D7-brane intersects the D3-branes at a fixed position on the $S^1$ of $x_3$, and wraps a $S^4$ of the $S^5$. Note that in this case, similar to the D4/D6 case \cite{Kruczenski:2003uq}, a single D7-brane with constant $x_3$ along the radius $r$ terminating at $r_t$ is not a physical solution. Instead, one should consider a pair of D7- and $\overline{D7}$-branes connecting at $r_t$ and ending on antipodal points of the $S^1$, analogous to the Sakai-Sukimoto model \cite{Sakai:2004cn} of the D4/D8/$\overline{\rm D8}$ configuration.

The D7 wraps a $S^4$, accordingly the $S^5$ metric is written as
\be
d\Omega_5^2=d\chi^2+\sin^2\chi d\Omega_4^2.
\ee
The massless case corresponds to $\chi=\pi/2$ so that the probe brane wraps the maximal $S^4$ and the separation (proportional to $\cos\chi$) between the D7- and D3-branes in their common transverse space is zero.

 In the static gauge, the induced metric on the D7 brane is
\be
ds_{D7}^2=\frac{r^2}{R^2}(-dt^2+dx_1^2+d x_2^2)+\frac{R^2dr^2}{fr^2}+R^2 d\Omega_4^2.
\ee
The DBI action yields
\be
\cl=-T_7 V_{S^4} R^2\int_{r_t}^\infty dr \frac{r^2}{\sqrt{f}}\sqrt{1-\frac{r_*^4}{r^4}},
\ee
where $\cl=S_{DBI}/V_3$,  the volume of the unit $S^4$ is $V_{S^4}= \frac{8}{3}\pi^2$, and $r_*$ is defined as before. The imaginary part is
\be
{\rm Im}\cl=T_7V_{S^4} R^2\int_{r_t}^{r_*}dr r^2\frac{\sqrt{r_*^4-r^4}}{\sqrt{r^4-r_t^4}}=T_7V_{S^4} R^2\frac{r_*^3}{3}\int^1_{y_t} dy\frac{\sqrt{1-y^{4/3}}}{\sqrt{y^{4/3}-y_t^{4/3}}},
\ee
where $y\equiv r^3/r_*^3<1$ and $y_t\equiv r_t^3/r_*^3<1$.

The integral can be evaluated and expressed in terms of hypergeometric functions, whose particular form   does not concern us. The key point is that one can easily find exactly the same relations \eq{critical E} and \eq{comparison Mkk}. Indeed, these are properties of the D3-branes background, which should be independent of the configurations of the probe branes. Note, however, that in this case ${\rm Im}\cl\propto r_*^3\propto E^{3/2}$, a feature of a 2+1 dimensional theory. This is, of course, not due to the KK mechanism; indeed, the energy scale corresponding to $E_c$ is higher than the KK scale. The reason is that the D7-brane intersects the $S^1$ wrapped by the D3-branes, leading to a 2+1 defect within the D3-brane world volume and the Schwinger effect occurs only in the this restricted region.

\section{Scwhinger effect in holographic NCSYM}

\subsection{Properties of NCSYM and its holographic dual}
In NCSYM, the ordinary SYM theory is generalized to noncommutative (NC) spacetime, where the NC coordinates obey the Heisenberg algebra $[x^i,y^i]=i\theta^{ij}$, and products of fields are replaced by the $\star$-product, e.g. for   NC coordinates $x_2$ and $x_3$,
    \be
    f\star g(x_2,x_3)\equiv e^{\frac{i}{2}\theta(\partial_2\partial_3-\partial_3\partial_2)} f(x_2,x_3)g(x_2,x_3),
    \ee
    where the position  of the derivative operators is in accordance with the position of the function $f$ and $g$.

The holographic description of a NCSYM at large $N$ and strong 't Hooft coupling is given by \cite{Hashimoto:1999ut,Maldacena:1999mh}
\be
ds^2=u^2\left [-fdt^2+dx_1^2+h(u) (dx_2^2+dx_3^2)\right]+\frac{du^2}{u^2 f}+d\Omega_5^2,
\ee
with nonconstant dilaton and nontrivial $B$ field
\be
e^{2\phi}=g^2_s h(u),\ \ \ \ B_{23}=a^2 u^4 h(u),
\ee
where $f=1-u_h^4/u^4$, $h(u)=1/(1+a^4 u^4)$, and $a=\lambda^{1/4}\sqrt{\theta}$. The NC property arises from the presence of the NS $B$ field, whose only nonvanishing component in this case is $B_{23}$. $\theta$ is the weak coupling NC parameter in the NCSYM, $a$ can be regarded as the `renormalized'  NC parameter which contains quantum corrections \cite{Hashimoto:1999ut}. Note $R^4=4\pi g_s N_c\alpha'^2 $ has been set to unity for convenience.

In the IR limit  $u\rightarrow u_h$, as long as $u_h a\ll1$, $h$ becomes unity and the metric becomes a Schwarzschild black hole in $AdS_5\times S^5$.     Note that $u$ plays the role of an energy scale in the SYM. Thus the condition $u_h a\ll1$ can be understood as
        that the energy $u_h$ is much less than the UV scale $a^{-1}$ where the NC effect becomes significant, or equivalently,
        $T^{-1}\gg a$ ($T\propto u_h$), i.e. the thermal length is much larger than the NC length.

In the UV limit $u\rightarrow   \infty$, $g_{22}=g_{33}\propto h=0$, which seems to indicate that the metric of the dual theory on the boundary becomes degenerate. This is not the case.
In fact, in the presence of a bulk NS  $B$ field, the metric relevant to the boundary field theory is {\it not} the closed string metric $g_{ij}$, but the open string metric \cite{Seiberg:1999vs,Li:1999am} $G_{ij}=g_{ij}-(B g^{-1} B)_{ij}$.
Then the spatial metric of the NCSYM becomes the usual flat one $\delta_{ij}$

\subsection{Calculation of the decay rate}
Now consider the action of a probe D7 brane
in the above D3-branes background with the $B$ field.
Again, only the DBI part of the probe brane action is relevant since we are concerned with the imaginary part,
\be
 S_{DBI}=-T_7\int d^8x  e^{-\phi}\sqrt{-\det(G_{ab}+B_{ab}+2\pi\alpha' F_{ab})},
\ee
where  $B_{ab}$ is the pullback of  the $B$ field on the probe brane, and we put down the dilaton explicitly since it is not a constant now. In the following we denote the gauge field components as
\be
E_1=F_{01},\ \ \ B_i=\frac{1}{2}\epsilon_{ijk}F_{jk},
\ee
where $\epsilon_{ijk}$ is the Levi-Civita symbol.

As before, we still consider the massless case, where the D7-brane wraps the maximal $S^3$ of the $S^5$. That this flat hyperplane is still a solution to the embedding equation  can be seen as follows. In the presence of the $B$ field, the DBI part is the same as that without the $B$ field, while the Chern-Simons part is zero. Thus the embedding equation arising from variation of the probe brane action is the same as the one in the usual D3/D7 case without the $B$ field, in which case the flat hyperplane is a solution, as has already been used in \cite{Hashimoto:2013mua}. Then, using the static gauge, the   result takes the form
\be
    S_{DBI}=-2\pi^2 T_7 g_s^{-1}V_4\int_{u_h}^\infty du \frac{u^3}{\sqrt{h}}\sqrt{\xi}.
\ee


Given that the rotational symmetry in the three space is partially broken by the $B$ field, it is naively expected that the physics should be different for   two typical orientations of the electric field, i.e.
(1) $E$  in the ordinary direction $x_1$ and (2) $E$ in one of the NC direction, e.g. $x_2$. As it turns out, however, calculations in the two cases  in the absence of the magnetic field give rise to the same result,
    \be
    \xi=h(1-\frac{E^2\alpha^2}{u^4f }).
    \ee
    where, for convenience, we  denote $\alpha\equiv2\pi\alpha'$.
    The $h$-dependence, which characterizes the NC effect, cancel out in $S_{DBI}$.  Consequently, one obtains  exactly the same result of the D3/D7 without the $B$ field dual to ordinary SYM \cite{Hashimoto:2013mua}. So the decay rate is not affected by the NC effect.

Furthermore, one can consider the cases when the magnetic field on the probe brane is turned on:
\bnum
    \item
    $E$ in the ordinary direction $x_1$, $\vec E\rightarrow E_1$. Then $B_2$ and $B_3$ are normal while $B_1$ is parallel to $\vec E$.
    \be
    \xi=h+2a^2h\alpha B_1(1-\frac{\alpha^2 E_1^2}{u^4f})-\frac{\alpha^2}{u^4}[E_1^2 f^{-1}h-(B_1^2+(B_2^2+B_3^2)h)]-\frac{\alpha^4E_1^2 B_1^2}{u^8f}.
    \ee

    \item
    $E$ in the NC direction $x_2$, $\vec E\rightarrow E_2$. Then $B_1$ and $B_3$ are normal while $B_2$ is parallel to $\vec E$.
    \be
    \xi=h+2a^2h\alpha B_1-\frac{\alpha^2}{u^4}[E_2^2 f^{-1}h-(B_1^2+(B_2^2+B_3^2)h)]-\frac{\alpha^4 E_2^2B_2^2}{u^8 f}.
    \ee

\enum

Comments
\bi
    \item
    When we turn off the NC effect by setting $a=0$, i.e. $h=1$, the above two equations coincide and essentially recover the result of the DBI action in the ordinary D3/D7 system without the $B$ field in \cite{Hashimoto:2014dza} \footnote{In particular, their equations (2.17) or (2.19)}.
 \item
In both cases, when $B_1=0$, i.e. no $F_{23}$ is added to the nonvanishing component of the $B$ field in the DBI action, the above results recover  exactly the ones in the ordinary D3/D7 case i.e. Eq.(2.19) of \cite{Hashimoto:2014dza}.
In other words, the NC effect affects the DBI action and the decay rate only when the magnetic component $B_1=F_{23}$ is nonvanishing.

 \item
 In the zero temperature limit, $u_h\rightarrow0$, $f\rightarrow 1$. The dominant contribution to the imaginary part as $u\rightarrow 0$ comes from the last term in $\xi$, which is not modified by the NC effect. Thus the same conclusion holds as in \cite{Hashimoto:2014dza} that there is a logarithmic divergence in the decay rate at zero temperature, and the small temperature limit is ${\rm Im}\cl\sim -\frac{N_c}{4\pi^2}\vec E\cdot \vec B\log T$.

\ei

\section{Conclusion}
By evaluating the imaginary part of the DBI action of the probe brane, we obtained the flavor pair production rate in large $N$ strongly coupled 3+1 dimensional SYM compactified on a circle. The confining D3-brane background has been studied in the potential analysis \cite{Sato:2013dwa}. Moreover, in \cite{Hashimoto:2014yya}, the decay rate was calculated in a similar confining background, i.e. the Sakai-Sugimoto model of holographic QCD \cite{Sakai:2004cn}. Something new in our work is that we emphasize that in the confining D3-brane background, the Schwinger effect occurs at an energy scale higher than the KK mass, where the tower of KK massive modes contribute and the theory is not an effective 2+1 dimensional one anymore.

Applying the same method to the holographic NCSYM, we found the way in which the NC effect modifies the decay rate. In particular, the NC effect brings about no change  to the decay rate once the magnetic component $F_{23}$ is absent. When such component is turned on, the NS $B$ field effectively acquires a constant shift. This makes it impossible to cancel the NC effect, i.e. the $h$-dependence, in the DBI action\footnote{Similar effect can be also found in \cite{Seo:2009um}.}.

In the large $N$ limit, physics of NCSYM at weak coupling is argued to be the same as ordinary SYM \cite{Bigatti:1999iz}. Moreover, when a supergravity dual description is valid, studies of thermodynamics \cite{Maldacena:1999mh, Cai:1999aw,Cai:2000hn} of the D3-branes background with a nonvanishing $B$ field using a probe D3-brane lends further support to such equivalence between NCSYM and ordinary SYM in the large $N$ limit at strong coupling. Our result here seems to indicate some inequivalence when it comes to the Schwinger effect. It would be interesting to further understand the NC effect on the Schwinger effect.

\acknowledgments

The author would like to thank Rong-Gen Cai and Qing Yang for invaluable discussions.  This work was supported in part by the project of Knowledge Innovation Program of Chinese
Academy of Science, NSFC under Grant No.11175225, No.11375247
and National Basic Research Program of China under
Grant No.2010CB832805.

\bibliographystyle{JHEP}
\bibliography{Biblio}
\end{document}